\documentclass[prl,aps,showpacs,byrevtex,onecolumn]{revtex4}
\usepackage{amssymb}
\usepackage{graphicx}
\usepackage{amsmath}
\usepackage{amsfonts}

\begin{document}

\title{Inverse problem and Bertrand's theorem }
\author{Yves Grandati$^1$, Alain B\'erard$^1$, and Ferhat M\'enas$^2$}
\address{$^1$Laboratoire de Physique Moléculaire et des Collisions, ICPMB, IF CNRS 2843,
Universit\'e Paul Verlaine, Institut de Physique, Bd Arago, 57078 Metz, Cedex 3, France \\
$^2$Laboratoire de Physique et de Chimie Quantique, Facult\'e des Sciences, Universit\'e Mouloud Mammeri,  BP 17 Tizi Ouzou, Algerie}

\date{\today}

\begin{abstract}
The Bertrand's theorem can be formulated as the solution of an inverse
problem for a classical unidimensional motion. We show that the solutions of
these problems, if restricted to a given class, can be obtained by solving a
numerical equation. This permit a particulary compact and elegant proof of
Bertrand's theorem.
\end{abstract}


\maketitle

\subsection{Introduction}

In classical point mechanics, since central potentials are isotropic they're
at the basis of every two body interactions. Among them we find in
particular Hooke's and Newton's potentials which possess very specific
properties. A part of them were already known by these pionners of modern
theoretical physics. For example, they're the only potentials presenting
elliptic bound states'orbits and the ellipses associated to each of them are
in dual relation\cite{arnvas}.

The question to determine if others central potentials could
generate closed bound states orbits for every values of the
initial parameters of the motion (energy and angular momentum)
stayed opened for almost two centuries. In 1873, J.\ Bertrand has
shown that the answer to this question is negative. Since thirty
years, a number of others proofs of this fundamental result have
been proposed. With an exception \cite{salas},\cite{martinez},
every proof schemes can be decomposed in three steps. The first
one establish that for the researched potentials, the angular
period, also called apsidal angle, is necessarily independent of
energy and angular momentum. In the second step, one shows that
constancy with respect to angular momentum, when applied to orbits
closed to circular ones, leads to retain only power law
potentials. Both of these steps are common to every proofs. Only
the last step is handled differently. It consists to show that
only Hooke's and Newton's potentials lead always to closed orbits.
Most of the proofs use a perturbative approach \cite{brown}
,\cite{gold},\cite{fejoz},\cite{zarmi}, \cite{tikoch}. The
original proof \cite{bert},\cite{greenb} and the inspired ones
\cite{arnold} follow rather a global approach. Since the problem
is to determine a class of potentials from informations about the
orbital angular period, the most natural and direct approach is to
interpret it as an inverse problem. This approach has already been
used by Y.\ Tikochinsky \cite {tikoch} and in a different and more
general setting by E. Onofri and M. Pauri \cite{onofri}. In
Tikochinsky's work \cite{tikoch}, it effectively runs into a
simplified proof, but the inverse problem treatment still
necessitates tedious perturbative calculations.

In this paper, we propose to treat the inverse problem from a new point of
view, then obtaining a particularly compact and elegant proof of Bertrand's
theorem. Moreover, we will show that this formulation permits as well to
deal with the problem perturbatively.

\subsection{Basics about the motion of a particle in a
central force field}

Let's consider a particle submitted to a central force field
deriving from the potential $U(r)$. The angular momentum $\overrightarrow{L}%
\left( t\right) =\overrightarrow{p}_{0}\times \overrightarrow{r}_{0}$ is
then conserved and every orbit lies in the plane perpendicular to $%
\overrightarrow{L}$.

In polar coordinates $\left( r,\varphi \right) $, the equation of motion in
this plane gives \cite{arnold},\cite{gold},\cite{land},\cite{Whittaker} :
\begin{equation}
\left\{
\begin{array}{c}
\stackrel{..}{r}(t)+\frac{1}{m}V_{L}^{\prime }(r(t))=0 \\
\stackrel{.}{\varphi }\left( t\right) =\frac{L}{mr^{2}\left( t\right) }
\end{array}
\right.  \label{eqrad}
\end{equation}
where $V_{L}(r)$ is the \textbf{radial effective potential}, that is the sum
of the initial potential and the centrifugal barrier term $\frac{L^{2}}{%
2mr^{2}}$ which intensity depends of the angular momentum :
\begin{equation}
V_{L}(r)=U(r)+\frac{L^{2}}{2mr^{2}}
\end{equation}
The radial coordinate $r(t)$ describes an autonomous unidimensionnal motion,
those of a particle submitted to $V_{L}(r)$.
\begin{figure}
\includegraphics{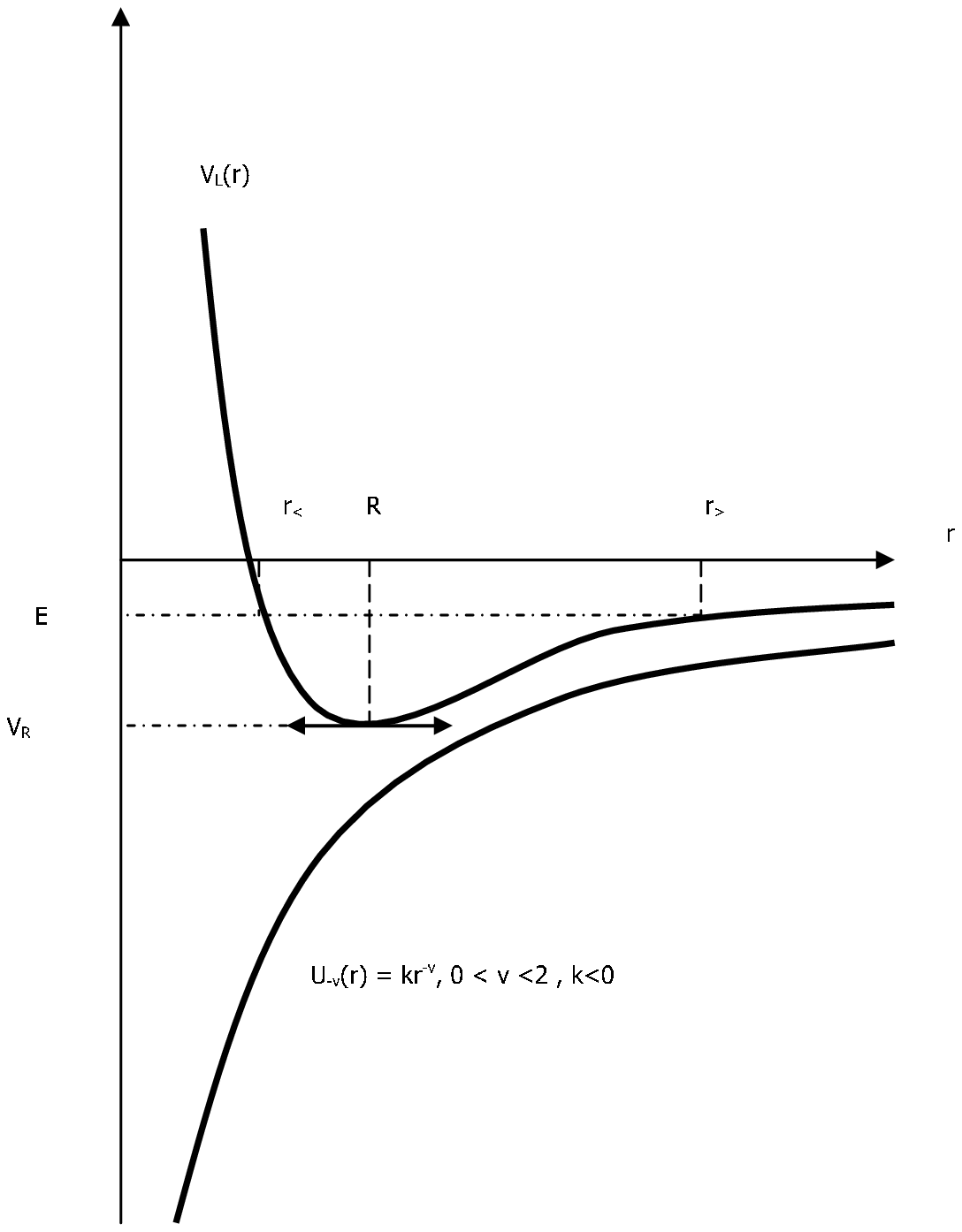}
\caption{Radial effective potentials for power law potentials}
\end{figure}

\begin{figure}
\includegraphics{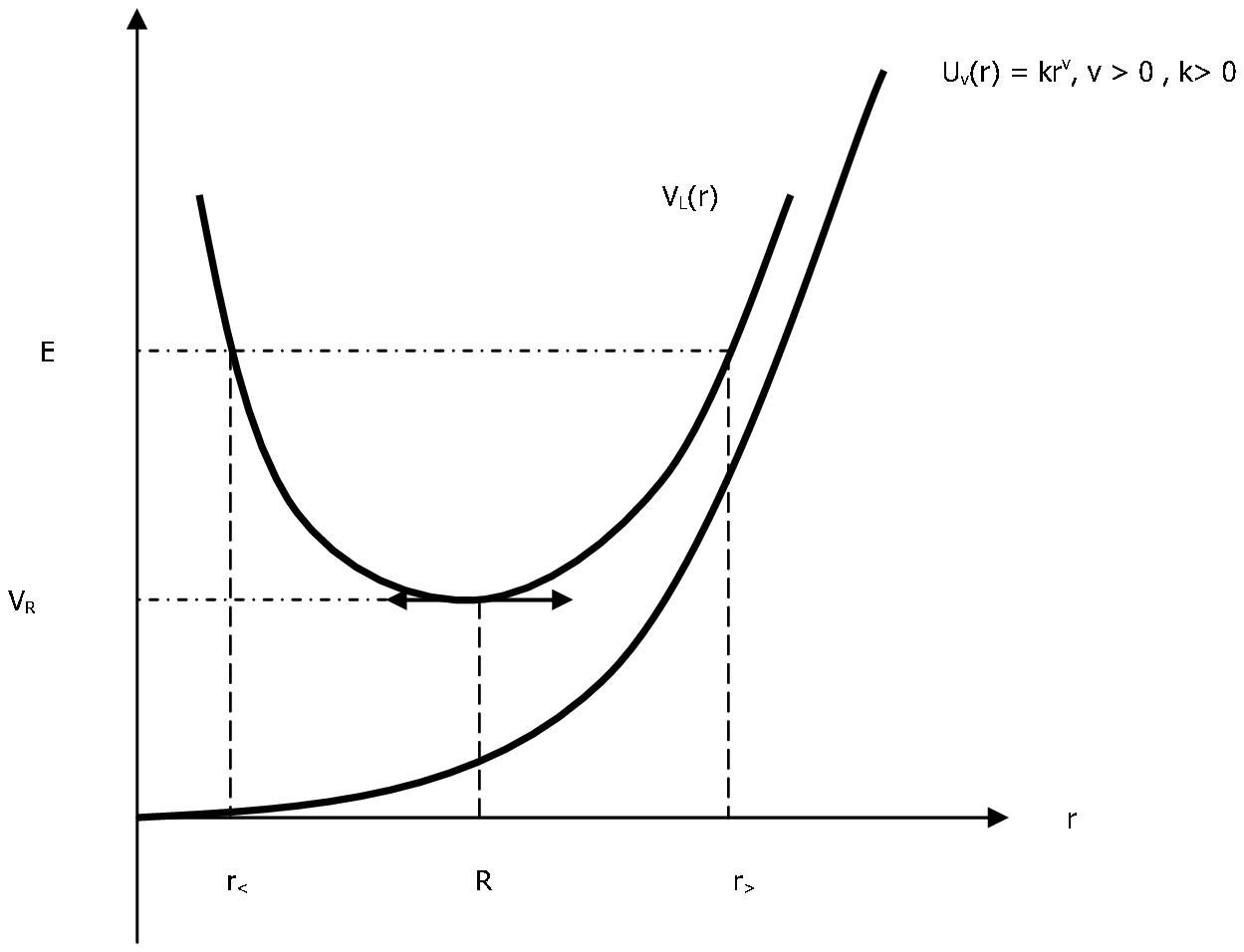}
\caption{Radial effective potentials for power law potentials}
\end{figure}

The solution $r(t)$ of equation (\ref{eqrad}) is given, at least
implicitly, by Barrow's formula :
\begin{equation}
t=\left( \frac{m}{2}\right) ^{\frac{1}{2}}\int_{r_{0}}^{r(t)}\frac{d\rho }{%
\sqrt{E-V_{L}(\rho )}}
\end{equation}
where $E$ is the conserved energy of the system and where we have chosen the
initial conditions $t_{0}=0,\ r(0)=r_{0},\ \varphi (0)=\varphi _{0}$.

The angular coordinates $\varphi \left( t\right) $ is obtained from $r(t)$
by a simple integration :
\begin{equation}
\varphi \left( t\right) -\varphi _{0}=\frac{L}{m}\int_{0}^{t}\frac{dt}{%
r^{2}(t)}
\end{equation}
We therefore lead to a complete parametric description $\left( r(t),\varphi
(t)\right) $ of the motion with respect to the time.

Let's introduce some elements of the vocabulary usually used for
the description of this type of motion.

* $r_{m}=\underset{i}{\inf} r_{i}$ such as $E=V_{L}(r_{i})$ is called a
\textbf{pericentral radius} and we have $E<V_{L}(r_{m}-\varepsilon )$,\ $%
\forall \ \varepsilon >0$ sufficiently small. Every point $A_{m}$ such that $%
\left\| \overrightarrow{OA_{m}}\right\| =r_{m}$, is a \textbf{pericenter}.

* $r_{M}=\underset{i}{\sup }r_{i}$ such as $E=V_{L}(r_{i})$ is called a
\textbf{apocentral radius} and we have $E<V_{L}(r_{M}+\varepsilon )$,\ $%
\forall \ \varepsilon >0$ sufficiently small. Every point $A_{M}$ such that$%
\left\| \overrightarrow{OA_{M}}\right\| =r_{M}$ is an \textbf{apocenter}.

* $r_{m}$ and $r_{M}$ are the \textbf{apsidal distances }of the motion the
vector positions $\overrightarrow{r}_{A_{m}}$et $\overrightarrow{r}_{A_{M}}$
of p\'{e}ricenters and apocenters are called \textbf{\ apsidal vectors}. The
angle $\Phi $ between two consecutive apsidal vectors is the\textbf{\
apsidal angle}.

If $r_{M}<+\infty $ , the orbit is \textbf{bounded}. If therefore $r_{m}>0$,
the radial motion is an oscillatory motion between the two extremal values $%
r_{m}$ and $r_{M}$. The orbit is then localized in the annulus
$r_{m}\leq r\leq r_{M}$. Such a bounded orbit will only be closed
at the condition that $2\Phi $ (which is the angle between two
consecutive pericenters or two consecutive apocenters) be
commensurable with $2\pi $, that is :
\[
\Phi =\frac{p}{q}\pi \in \pi \Bbb{Q}\quad
\]
where $p$ and $q$ are integers. In this case indeed, after $p$ revolution
around the origin, the particle makes $q$ radial oscillations.

In every others cases, that is $\Phi \notin \Bbb{Q\pi }$, the bounded orbit
is everywhere dense in the annulus $r_{m}\leq r\leq r_{M}$. This kind of
orbit is called a \textbf{rosette}.

\subsubsection{Circular orbits}
If the orbit is bounded there exists at least one absolute minimum for $%
V_{L}(r)$ on the interval $\left[ r_{m},r_{M}\right] $. We'll note
$R$ the corresponding value of the radial abscises and :
\begin{equation}
V_{R}=V_{L}(R)=U(R)+\frac{L^{2}}{2mR^{2}}
\end{equation}
The minimum conditions give :
\begin{equation}
\left\{
\begin{array}{c}
V_{L}^{\prime }(R)=0\Rightarrow \frac{L^{2}}{m}=R^{3}U^{\prime }(R) \\
V_{L}^{\prime \prime }(R)=\frac{RU^{\prime \prime }(R)+3U^{\prime }(R)}{R}%
\geq 0
\end{array}
\right.  \label{min}
\end{equation}
If $E=V_{R}$, the only authorized value for $r$ is $R$ and : $%
r(t)=R=r_{m}=r_{M},\ \forall t$ with $r(\varphi )=R,\ \forall \varphi $,
which corresponds to a \textbf{\ circular orbit} with radius $R$.

Along such a circular orbit we have :
\begin{equation}
\left\{
\begin{array}{c}
r(t)=R \\
\varphi (t)=\omega t+\phi _{0}
\end{array}
\right.
\end{equation}
If the initial potential $U(r)$ is such that the function $%
f(r)=r^{3}U^{\prime }(r)$, is locally bijective around each minimum of $%
V_{L}(r)$, the first of the\ above conditions (\ref{min}) shows
that we can indifferently choose to characterize circular orbits
by their radius $R$ or by the associated angular momentum $L$.

Let's finally note that in order to admit a finite circular orbit, the power
law potential $U(r)\sim r^{\nu }$ must satisfy $\nu >0$ or $-2<\nu <0$, that
is $U(r)$ must be of the form :
\begin{equation}
\left\{ \begin{array}{c}
U_{\nu}(r)=kr^{\nu },\quad \nu >0 \\
U_{-\nu}(r)=-kr^{-\nu },\quad 0<\nu <2
\end{array}
\right. \quad k>0
\end{equation}
In each case we have a unique circular orbit with a radius respectively
equal to :
\begin{equation}
\left\{
\begin{array}{c}
R_{\nu}=\left( \frac{L^{2}}{\nu km}\right) ^{\frac{1}{\nu+2}},\quad \nu >0 \\
R_{-\nu}=\left( \frac{L^{2}}{\nu km}\right) ^{\frac{1}{-\nu+2}},\quad 0<\nu <2
\end{array}
\right.
\end{equation}

\subsection{Clairaut's variable and Binet's equation}
If our only ambition is to describe the orbit, it is possible to
obtain its polar equation in a direct way. To do it, it's
sufficient to note that, if we consider now $r$ as a function of
$\varphi $, \'{e}equation (\ref{eqrad})
becomes $\left( \frac{d}{dt}=\frac{L}{mr^{2}(\varphi )}\frac{d}{d\varphi }%
\right) $ :
\begin{equation}
\frac{d^{2}r(\varphi )}{d\varphi ^{2}}-\frac{2}{r(\varphi )}\left( \frac{%
dr(\varphi )}{d\varphi }\right) ^{2}+\frac{m^{2}r^{4}(\varphi )}{L^{2}}%
V_{L}^{\prime }(r(\varphi ))=0  \label{eqorb}
\end{equation}
Its solutions furnish directly the orbital equation under the form $%
r=r(\varphi )$. This equation is considerably simplified if we do the
following change of variable $x=\frac{L}{mr}$ (due to Clairaut \cite{Whittaker}).
\textbf{Clairaut's variable} $x$, is nothing else, upon to a
constant factor, that the inverse radial variable. $x$ and $\varphi $ are
called \textbf{Clairaut's coordinates}.

With this change of coordinate, equation (\ref{eqorb}) then becomes :
\begin{equation}
\frac{d^{2}x(\varphi )}{d\varphi ^{2}}+\frac{1}{m}W_{L}^{\prime }\left(
x(\varphi )\right) =0
\end{equation}
where :
\begin{equation}
W_{L}(x)=V_{L}\left( \frac{L}{mx}\right) =\frac{1}{2}mx^{2}+U\left( \frac{L}{%
mx}\right)  \label{potClairaut}
\end{equation}
This identity is more known as \textbf{Binet's formula or Binet-Clairaut's
equation}.

When $\varphi $ varies, Clairaut's variable $x$ then describes a
one-dimensional motion those of an effective particle of masse m
submitted to the \textbf{Binet-Clairaut's potential }$W_{L}(x)$.
The evolution parameter of this motion is the angular position
$\varphi $, growing with time. When $r(\varphi )$ makes an
oscillation between the values $r_{m}$ and $r_{M}$ then $x(\varphi
)$ makes a corresponding oscillation between the two extrema
$x_{>}=\frac{L}{mr_{m}}$ and $x_{<}=\frac{L}{mr_{M}}$. Let's note
that to each extremum of $V_{L}(r)$ in $r=R$ corresponds a
extremum of the same type for $W_{L}(x)$ in $x_{0}=\frac{L}{mR}$.
Concerning the curvature of the Clairaut's potential near a
minimum $x_{0}$, that is a circular orbit, it writes :
\begin{equation}
W^{\prime \prime }(x_{0})=m\frac{RU^{\prime \prime }\left( R\right)
+3U^{\prime }\left( R\right) }{U^{\prime }(R)}\geq 0  \label{apscirc}
\end{equation}

Choosing the angles origin at an apocentral vector $\overrightarrow{r}%
_{A_{M}}$ and taking the initial condition $x\left( 0\right) =x_{<}$,
Barrow's formula, when applied to Binet-Clairaut's equation, gives an
implicit solution for the orbital equation :
\begin{equation}
\varphi (x)=\left( \frac{m}{2}\right) ^{\frac{1}{2}}\int_{x_{<}}^{x}\frac{%
d\xi }{\sqrt{E-W_{L}(\xi )}},\quad \forall x\in \left[ x_{<},x_{>}\right]
\end{equation}
Then the apsidal angle $\Phi $, which is Clairaut's motion and the
half-period of the radial oscillation is :
\begin{equation}
\Phi (E,L)=\left( \frac{m}{2}\right) ^{\frac{1}{2}}\int_{x_{<}}^{x_{>}}\frac{%
dx}{\sqrt{E-W_{L}(x)}}  \label{anglaps1}
\end{equation}
This expression takes a very compact form if we use the semi-derivative's
concept\cite{grandati}. The \textbf{semi-derivative} is the\textbf{\ }%
integral operator $D_{E}^{\frac{1}{2}}$ defined as \cite{miller} :
\begin{equation}
D_{E}^{\frac{1}{2}}g(E)=\frac{1}{\sqrt{\pi }}\int_{V_{R}}^{E}dw\frac{1}{%
\sqrt{E-w}}g^{\prime }(w)  \label{semider}
\end{equation}
where the function $g(E)$ satisfies $g(V_{R})=0$.

The semi-derivative is a particular case of \textbf{\ fractional
derivative}, notion about which an abundant mathematical
literature is available \cite {miller},\cite{Oldham},\cite{samko}
and which has found today many physical applications
\cite{west},\cite{flores},\cite{grandati}.

Putting $\Delta x(w)=x_{>}\left( w\right) -x_{<}\left( w\right) $, where $%
x_{>}\left( w\right) $ and $x_{<}\left( w\right) $ are the reciprocals of $%
W_{L}(x)$ defined on each branches of this last, on both sides of $x_{0}$,
equation (\ref{anglaps1}) becomes :
\begin{equation}
\Phi (E,L)=\sqrt{\frac{m\pi }{2}}D_{E}^{\frac{1}{2}}\Delta x(E)
\label{anglaps2}
\end{equation}

If the apsidal angle's notion becomes meaningless for a strictly
circular orbit, we can however try to calculate the limit value it
takes when the considered orbit is in the neighborhood of a
circular one. From the point of view of the radial or Clairaut's
motions ($r(t)$ or $x(\varphi )$), this
corresponds to small oscillations near the equilibrium values $R$ and $x_{0}=%
\frac{L}{mR}$ of potentials $V_{L}(r)$ and $W_{L}(x)$ respectively.

Let's recall \cite{arnold},\cite{land} that in the small
oscillations limit, a particle whose potential $W(x)$ has a non
zero curvature in the vicinity of the equilibrium position $x_{0}$
is, at the first order of approximation an harmonic, (that is
isochronous) oscillator with frequency :
\begin{equation}
\omega =\sqrt{\frac{W^{\prime \prime }(x_{0})}{m}}
\end{equation}
In the present context, this simply gives (\ref{apscirc}) :
\begin{equation}
\Phi _{C}(E,L)=\pi \sqrt{\frac{U^{\prime }(R)}{RU^{\prime \prime
}(R)+3U^{\prime }(R)}}=\Phi _{C}(R)
\end{equation}
where \ the angular period of radial oscillations in the vicinity
of a circular orbit $\Phi _{C}(R)$, is now independent of the
energy $E$.

Knowing the function $\Phi _{C}(R)$, the above identity becomes a
second order linear differential equation for $U(R)$ whose
solution is readily obtained as :
\begin{equation}
U(r)=\int^{r}dr^{\prime }e^{-\int^{r^{\prime }}d\rho \frac{1}{\rho }\left(
3-\left( \frac{\pi }{\Phi _{C}(\rho )}\right) ^{2}\right) }
\end{equation}

In the following we will specially refer to the case where $\Phi _{C}(R)$ is
a constant $\Phi _{C}$, independent of the radius $R$ (that is of the
angular momentum $L$ of the particle). In this case, from the above formula
we obtain two distinct functional forms for $U(r)$ :
$\ast $ If $\Phi _{C}\neq \frac{\pi }{\sqrt{2}}$, defining the exponent $v$
as $\Phi _{C}=\frac{\pi }{\sqrt{2+\nu }}$, $\nu \neq 0$ :
\begin{equation}
U(r)=Ar^{\nu }+B  \label{powlerawpot}
\end{equation}
$\ast $ If $\Phi _{C}=\frac{\pi }{\sqrt{2}}$ (that is for $\nu =0$) :
\begin{equation}
U(r)=A\ln r+B  \label{logpot}
\end{equation}
Therefore, there are only two types of potentials for which the apsidal
angle $\Phi _{C}$ of a bounded orbit near a given circular one, is a
constant, ind\'{e}pendent of the characteristic parameter $L$ : power law
potentials and logarithmic potentials.

\subsection{Inverse problem for the Clairaut's potential}
The inverse problem for a classical one dimensional oscillator consists in
determining the oscillator potential from the variation law of the period as
a function of the total energy. Then to determine Clairaut's potential $%
W_{L}(x)$ from the variations of the apsidal angle $\Phi (E,L)$ as a
function of $E$ is a problem of this type. Fractional integro-differential
calculus as introduced above gives a very direct way to the solution \cite
{grandati}. Indeed, the semi-derivative operator $D^{\frac{1}{2}}$ ($\ref
{semider}$) admits, on the set of bounded functions near $V_{R}$, an inverse
$D^{-\frac{1}{2}}$ called \textbf{semi-integral} \cite{miller},\cite{grandati}\ and defined as :
\begin{equation}
D_{E}^{-\frac{1}{2}}g(E)=\frac{1}{\sqrt{\pi }}\int_{V_{R}}^{E}dw\frac{1}{%
\sqrt{E-w}}g(w)
\end{equation}
For our purpose, we'll more particularly note the following result for the
semi-integral of the constant 1 :
\begin{equation}
D_{w}^{-\frac{1}{2}}1=\frac{1}{\pi }\sqrt{w-V_{R}}  \label{fractder1}
\end{equation}
Since $\Delta x\left( w\right) $ satisfies the required condition, equation (%
\ref{anglaps2}) linking $\Delta x$ to $\Phi (E,L)$ is immediately inverted
and the inverse problem's solution for Clairaut's motion is simply given by
the implicit formula \cite{grandati} :
\begin{equation}
\Delta x\left( E\right) =x_{>}(E)-x_{<}(E)=\sqrt{\frac{2}{m\pi }}D_{E}^{-%
\frac{1}{2}}\Phi (E,L)  \label{deltax}
\end{equation}
Unfortunately, this equation yields only the difference
$x_{>}(E)-x_{<}(E)$ for a given $\Phi (E,L)$, which doesn't
generally allow to determine uniquely the two branches
$x_{>}\left( w\right) $ and $x_{<}\left( w\right) $\ of the
potential's reciprocal function. For a given $\Delta x\left(
w\right) $ there exists an infinity of possible multiform $x\left(
w\right) $. Therefore, if our goal is to precise what potential
function leading to a given $\Phi (E,L)$, we see that this problem
will generally admit an infinity of solutions \cite{land}.

To determine $W_{L}(x)$ uniquely it's necessary to add
supplementary constraints, that is to limit the research to a more
restricted class of potentials. For instance, if we consider only
potentials which are symmetrical with respect to the axis
$x=x_{0}$, we obtain :
\begin{equation}
x_{>}\left( w\right) =\frac{1}{\sqrt{2m\pi }}D_{w}^{-\frac{1}{2}}\Phi
(w,L)+x_{0}=2x_{0}-x_{<}(w)
\end{equation}
But even if we have obtained $x_{>}\left( w\right) $ and $x_{<}(w)$, we
still have to invert them to find an explicit form of $W_{L}(x)$. Usually
this step is not analytically tractable.

We can avoid this difficulty by translating the identity (\ref{deltax}) into
a functional equation for $W_{L}(x)$ \cite{grandati}. Indeed we have :
\begin{equation}
W_{L}(x)=W_{L}\left( x+\Delta x\right)
\end{equation}
in every point $x$ of the interval $I=\left] 0,x_{0}\right] $, when $\Delta
x\left( w\right) $ is calculated in $w=W_{L}(x)$.

Using (\ref{deltax}), we then obtain the following functional equation for $%
W_{L}(x)$ :
\begin{equation}
W_{L}(x)=W_{L}\left( x+\sqrt{\frac{2}{m\pi }}D_{w}^{-\frac{1}{2}}\Phi
(W_{L}(x),L)\right)
\end{equation}
In the special case where $\Phi (E,L)=\Phi _{C}(L)$ constant with respect to
$E$ (''isochronous motion''), this gives (\ref{fractder1}) :
\begin{equation}
W_{L}(x)=W_{L}\left( x+\alpha _{L}\sqrt{W_{L}(x)-V_{R}}\right) ,\quad
\forall x\in I  \label{eqfonct2}
\end{equation}
with $\alpha _{L}=\sqrt{\frac{2}{m}}\frac{2\Phi _{C}(L)}{\pi }$.

A priori this equation doesn't seem particularly simple to solve
analytically. However if we suppose we can limit our research to a
restricted class of potentials $W(x;\left\{ \nu_{i}\right\} )$
with a given functional form but depending upon $m$ parameters
$\left( \nu_{i}\right) _{i=1,...,m}$,
we are leading to a system of $n$ numerical equations for the $\nu_{i}$, with $%
n$ arbitrarily large :
\begin{equation}
\left\{
\begin{array}{c}
W(x_{1};\left\{ \nu_{i}\right\} )=W\left( x_{1}+\sqrt{\frac{2}{m\pi }}D_{w}^{-%
\frac{1}{2}}\Phi (W(x_{1};\left\{ \nu_{i}\right\} ),L);\left\{ \nu_{i}\right\}
\right) \\
... \\
W(x_{n};\left\{ \nu_{i}\right\} )=W\left( x_{n}+\sqrt{\frac{2}{m\pi }}D_{w}^{-%
\frac{1}{2}}\Phi (W(x_{n};\left\{ \nu_{i}\right\} ),L);\left\{ \nu_{i}\right\}
\right)
\end{array}
\right.
\end{equation}
where the $x_{j}$ are $n$ points on $I$.

Specially, for a family of isochronous potentials depending upon a unique
parameter $\nu$ ($m=1$), we obtain the following equation :
\begin{equation}
W(x_{1};\nu)=W\left( x_{1}+\alpha _{L}\sqrt{W_{L}(x_{1};\nu)-V_{R}}; \nu\right)
\label{eqfonct1}
\end{equation}
where $x_{1}$ is arbitrarily chosen on $I$ ($n=1$).

We can equally chose for $x_{1}$ the limit value $0^{+}$. With
regard to the possible behaviors of $W_{L}(x)$ at this point we'll
\ consider only the two following cases :

$*$ $W_{L}(x)$ tends to a finite value $W_{L,0}=W_{L}(0^{+})$ in $0$ (this
corresponds to the case where $U(r)$ tends to a finite value when $%
r\rightarrow \infty $). Equation (\ref{eqfonct1}) becomes :
\begin{equation}
W_{L,0}\left( \nu \right) =W_{L}\left( \alpha _{L}\sqrt{W_{L,0}\left( \nu \right)
-V_{R}};\nu \right)  \label{eqfcten0}
\end{equation}

$*$ $W_{L}(x)$ diverges as $C_{0}x^{-\mu _{0}}$ when $x\stackrel{>}{%
\rightarrow }0$ and as $C_{\infty }x^{\mu _{\infty }}$ when $x\rightarrow
\infty $ ($\mu _{0},\mu _{\infty },C_{0},C_{\infty }>0$ being a priori $v$
dependent). This corresponds to the case where $U(r)$ diverges as $r^{\mu
_{0}}$ when $r\rightarrow \infty $ and as $r^{-\mu _{\infty }}$ when $%
r\rightarrow 0$. Equation (\ref{eqfonct1}) leads then to the relation :
\begin{equation}
C_{0}x^{-\mu _{0}} \underset{x\stackrel{>}{\rightarrow }0}{\simeq }
C_{\infty }C_{0}^{\frac{\mu _{\infty }}{2}}\alpha _{L}^{\mu _{\infty }}x^{-%
\frac{\mu _{0}\mu _{\infty }}{2}}\Rightarrow \left\{
\begin{array}{c}
\mu _{\infty }=2 \\
1=C_{\infty }\alpha _{L}^{2}
\end{array}
\right.  \label{eqfctalinf}
\end{equation}


\subsection{The Bertrand's theorem}
All the preceding results permit to obtain a particularly compact proof of
Bertrand's theorem. This last establish that :

$\ast $ \textbf{The only potentials for which every orbits near a circular
one are closed independently of energy and angular momentum are the Newton
potential and the Hooke potential.}

If one except the proofs \cite{martinez},\cite{salas} based on the necessary
existence of supplementary constants of motion, all the other proof's
schemes can be decomposed in three steps, the two firsts being common to all
proofs.

\subsubsection{First step : Obtaining the variation law for the angular
period $\Phi $ as a function of $E$ and $L$}

The first step is the keystone of the proof. In order that all the bounded
orbits be closed it's necessary that on all the intervals of admissible
values for the parameters $L$ and $E$ the corresponding values of the
apsidal angle $\Phi $ are rational :
\[
\Phi (E,L)=\frac{p}{q}\pi \in \pi \Bbb{Q}
\]
that is belong to a discrete set.

$\Phi (E,L)$ being supposed to vary continuously with $L$ and $E$, the
fact that its image is contained in a discrete set implies that it's a constant with respect to $L$ and $E$.

If we note $\Phi _{C}$ the limit apsidal angle apsidal near the considered
circular orbit, then we necessarily have :
\begin{equation}
\Phi (E,L)=\Phi _{C}\in \pi \Bbb{Q}  \label{rationnalité}
\end{equation}
for all the considered values of $E$ and $L$.

This extremely constraining constancy condition is the source of
the result.

\subsubsection{Second step : Selection of the potentials for which the limit
apsidal angle near a circular orbit is not $L$ dependent}

The second step consists to use the preceding results (\ref{powlerawpot}), (\ref
{logpot}) concerning the potentials possessing circular orbits near which the
limit apsidal angle is $L$ independent. Among them, the logarithmic
potential $U_{\log }(r)=k\ln r+cste,\quad k>0$, can be immediately excluded
since it leads to a limit apsidal angle $\Phi _{C}=\frac{\pi }{\sqrt{2}}$
which is not a rational multiple of $\pi $ and then doesn't satisfy the
condition (\ref{rationnalité}).

Therefore, the only permitted potentials $U(r)$ are of the type $U_{\pm
\nu}(r)=\pm kr^{\pm \nu},\ k>0$, and their associated Clairaut's potentials
writes (\ref{potClairaut}) :
\begin{equation}
W_{\pm \nu}(x)=\frac{1}{2}m\left( x^{2}\pm A_{\pm }x^{\mp \nu }\right) ,\quad
A_{\pm }=2\frac{k}{m}\left( \frac{L}{m}\right) ^{\pm \nu}  \label{potClairaut2}
\end{equation}
with $\nu>0$ for the $+$ sign and $0<\nu<2$ for the $-$ sign.

In both cases, one has a unique circular orbit for which the limit apsidal
angle is (\ref{powlerawpot}) :
\begin{equation}
\Phi _{C}=\frac{\pi }{\sqrt{2\pm \nu }}
\end{equation}

\subsubsection{Third step : Determination of the admissible
power law potentials} Now it remains finally to select among all
the possible values of the exponent $\nu$, those permitting to
satisfy the required conditions. The differences between the
various proof's scheme appear at this level. The great majority
lie on a local study in the circular orbit's neighborhood, the
corrections being determined by an adapted perturbative approach
\cite {tikoch},\cite{gold},\cite{brown},\cite{zarmi},\cite{fejoz}.
The original proof \cite{bert}, \cite{greenb} and Arnold's one
\cite{arnold} proceed from a global approach. Now, we are going to
see that our inverse problem's formulation as developed in the
preceding paragraphs permits to solve the asked problem in a very
simple and direct way if we consider globally the potential
behavior. We will show that this formulation permits equally a
local perturbative treatment, requiring however tedious
calculations, which is a common feature of this kind of approach.

Let's come back to the asked problem. We have to determine all the
(Clairaut's) potentials leading to a given variation law for the
(angular) period as a function of energy. As noted by Tikochinsky,
we are typically in the frame of an inverse problem \cite{tikoch}.
Since in our case, the variation law is a constant, we are
therefore leaded to determine among all the Clairaut's potentials
in the families $\left( W_{\nu}(x)\right) _{\nu>0}$ et $\left(
W_{-\nu}(x)\right) _{0<\nu<2}$\ (\ref{potClairaut2})\ those being
isochronous, that is satisfying the functional equation
(\ref{eqfonct2}).

\paragraph{ Global approach}
As seen previously, by choosing an adapted value for $x_{1}$, the
functional equation (\ref{eqfonct2}) results in numerical equation
for $\nu$ (\ref {eqfonct1}). Let's take for $x_{1}$ the limit
value $0^{+}$. The behaviors at the origin of the potentials
$W_{\nu}(x)$ and $W_{-\nu}(x)$  are :
\begin{equation}
\left\{
\begin{array}{c}
W_{\nu}(x)\underset{x\stackrel{>}{\rightarrow }0}{\simeq }\frac{1}{2}%
mAx^{-\nu }\underset{x\stackrel{>}{\rightarrow }0}{\rightarrow }+\infty \\
W_{-\nu}(x)\underset{x\stackrel{>}{\rightarrow }0}{\rightarrow }W_{L,0}=0
\end{array}
\right.
\end{equation}
$*$ For $\left( W_{-\nu}(x)\right) _{0<\nu<2}$, equation (\ref{eqfcten0}) gives :
\begin{equation}
W_{-\nu}\left( \alpha \sqrt{\left| V_{R}\right| }\right) =k\left( \frac{L^{2}}{%
\nu km}\right) ^{\frac{\nu}{\nu-2}}\left( 2-\left( \frac{\nu}{4}\right) ^{-\frac{\nu}{2}%
}\right) =0
\end{equation}
where in the present case $\alpha =\sqrt{\frac{2}{m}}\frac{2\Phi _{C}}{\pi }=%
\sqrt{\frac{2}{m}}\frac{2}{\sqrt{2\pm \nu }}$.

We then obtain a transcendental equation for $\nu$ :
\begin{equation}
e^{-\frac{\nu}{2}\ln \frac{\nu}{4}}=2
\end{equation}
whose solutions are readily obtained as $\nu=1$\ and $\nu=2$.

Since $0<\nu<2$, only $\nu=1$ is an admissible value. This corresponds to an
initial potential $U(r)=-\frac{k}{r}$, that is the Newton's potential.

$*$ For $\left( W_{\nu}(x)\right) _{\nu>0}$, identity (\ref{eqfctalinf})\ takes
the form :
\begin{equation}
1=C_{\infty }\alpha ^{2}
\end{equation}
with $C_{\infty }=\frac{1}{2}m$ and $\alpha =2\sqrt{\frac{2}{m(2+\nu)}}$. Then
:
\[
\frac{4}{2+\nu}=1\Rightarrow \nu=2
\]
In this case, the initial potential $U(r)$ is harmonic : $U(r)=kr^{2}$.

To summarize, \textit{the only potentials satisfying the required conditions
of the Bertrand's theorem are the Newton potential and the Hooke potential,
which achieves our proof of the theorem}.

\paragraph{ Perturbative approach}
As mentioned previously, it's possible to recover this result via a
perturbative resolution of the functional equation (\ref{eqfonct2})
satisfied by Clairaut's potential.

For that purpose, let's introduce the left and right lateral displacements $%
\varepsilon _{-}$ and $\varepsilon _{+}$. They measure the respective
distances between $x_{0}$ and the two branches of the potential function $%
W_{L}(x)$ :
\begin{equation}
\left\{
\begin{array}{c}
\varepsilon _{-}=x_{0}-x \\
\varepsilon _{+}=\Delta x-\varepsilon _{-}
\end{array}
\right.
\end{equation}
Then, the functional equation satisfied by $W_{L}(x)$ is in every point $%
x<x_{0}$ :
\begin{equation}
W_{L}(x_{0}-\varepsilon _{-})-V_{R}=W_{L}\left( x_{0}+\varepsilon
_{+}\right) -V_{R}  \label{eqfonct3}
\end{equation}

\begin{figure}
\includegraphics{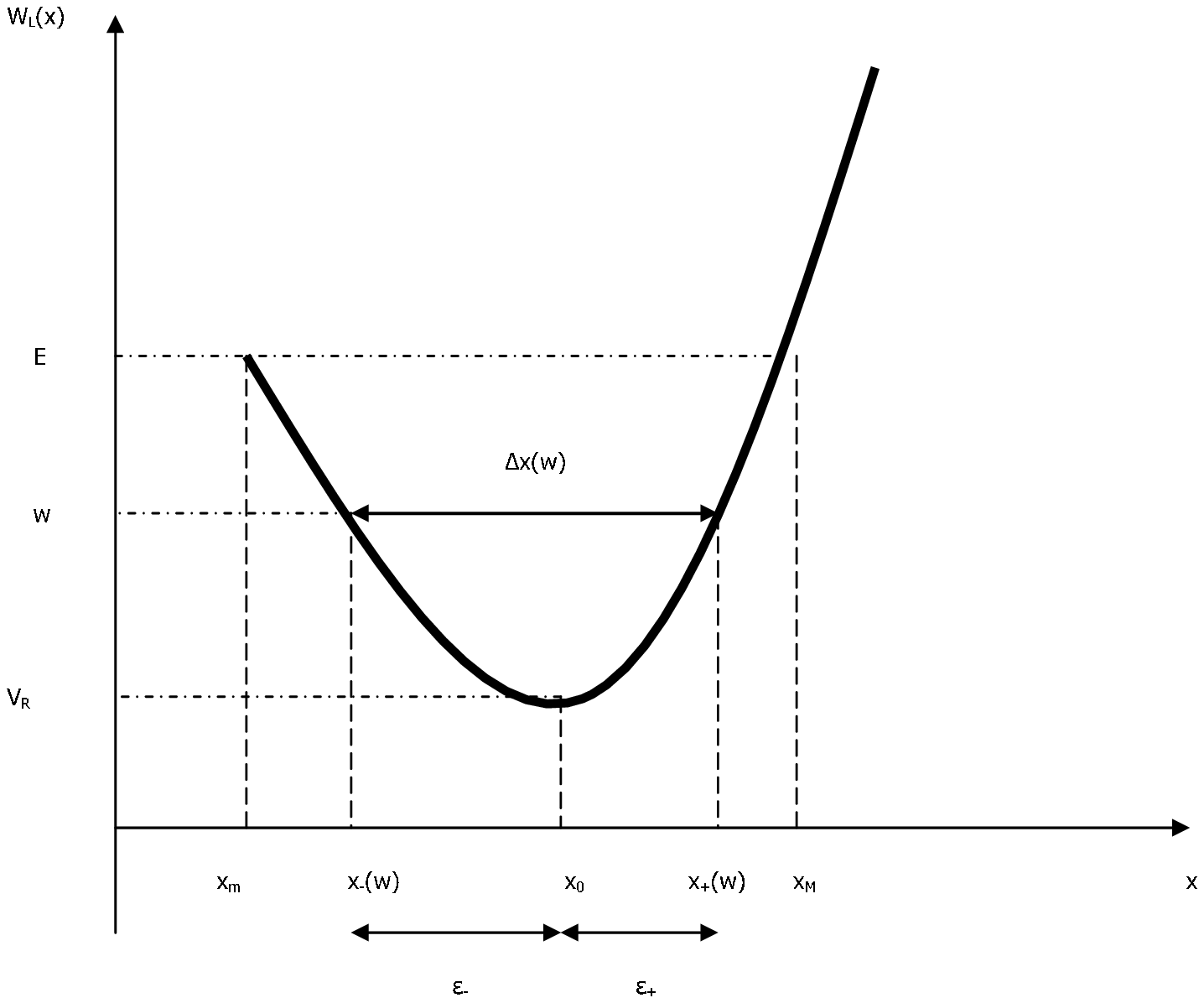}
\caption{Lateral displacements }
\end{figure}

If we consider an isochronous Clairaut's potential $W_{L}(x)$, the relation
between the lateral displacements writes :
\begin{equation}
\varepsilon _{+}=\Delta x-\varepsilon _{-}=\alpha \sqrt{W_{L}(x_{0}-%
\varepsilon _{-})-V_{L,m}}-\varepsilon _{-}  \label{rel entre depl lat}
\end{equation}
where $\alpha =\sqrt{\frac{2}{m}}\frac{2\Phi _{C}}{\pi }$.

For small amplitudes displacements, we can expand the left side of the above
identity in power of $\varepsilon _{-}$ .

First, we have :
\begin{equation}
W_{L}(x_{0}-\varepsilon _{-})-V_{R}=\frac{1}{2}m\omega ^{2}\varepsilon
_{-}^{2}\left( \sum_{n=0}^{\infty }\left( -1\right) ^{n}a_{n}\varepsilon
_{-}^{n}\right)
\end{equation}
where $W_{L}^{\prime \prime }(x_{0})=m\omega ^{2}$ and :
\begin{equation}
a_{n}=\frac{1}{\left( n+2\right) !}\frac{2W_{L}^{\left( n+2\right) }(x_{0})}{%
m\omega ^{2}}
\end{equation}
Inserting this expansion in the relation (\ref{rel entre depl lat}) between
lateral displacements, we obtain :
\begin{eqnarray}
\varepsilon _{+} &=&\varepsilon _{-}\frac{2\omega \Phi _{C}}{\pi }\sqrt{%
1-a_{1}\varepsilon _{-}+a_{2}\varepsilon _{-}^{2}+...+\left( -1\right)
^{n}a_{n}\varepsilon _{-}^{n}+...}-\varepsilon _{-} \\
&=&\varepsilon _{-}\left( \frac{2\omega \Phi _{C}}{\pi }-1\right)  \nonumber
\\
&&+\frac{2\omega \Phi _{C}}{\pi }\sum_{n=1}^{\infty }\frac{\left(
2n-3\right) !!}{2^{n}n!}\varepsilon _{-}^{n+1}\left( -a_{1}+a_{2}\varepsilon
_{-}+...+\left( -1\right) ^{p}a_{p}\varepsilon _{-}^{p-1}+...\right) ^{n}
\end{eqnarray}
that is, up to the fourth order in $\varepsilon $ :
\begin{equation}
\varepsilon _{+}=\varepsilon _{-}\left( 2\gamma -1\right) -\varepsilon
_{-}^{2}\gamma a_{1}+\varepsilon _{-}^{3}\gamma \left( a_{2}-\frac{a_{1}^{2}%
}{4}\right) +\varepsilon _{-}^{4}\gamma \left( -a_{3}+\frac{a_{1}a_{2}}{2}-%
\frac{a_{1}^{3}}{8}\right) +O\left( \varepsilon _{-}^{5}\right)
\label{epsplus}
\end{equation}
with $\gamma =\frac{\omega \Phi _{C}}{\pi }$.

Then, if we expand both members of the functional equation (\ref
{eqfonct3}) in power series of $\varepsilon _{-}$ and $\varepsilon _{+}$
respectively, we find :
\begin{equation}
\varepsilon _{-}^{2}\left( 1-a_{1}\varepsilon _{-}+a_{2}\varepsilon
_{-}^{2}+...+\left( -1\right) ^{n}a_{n}\varepsilon _{-}^{n}+...\right)
=\varepsilon _{+}^{2}\left( 1+a_{1}\varepsilon _{+}+a_{2}\varepsilon
_{+}^{2}+...+a_{n}\varepsilon _{+}^{n}+...\right)
\end{equation}
Inserting in this identity the above expression of $\varepsilon _{+}$ \ref
{epsplus}, we arrive at the formula :
\begin{eqnarray*}
1-a_{1}\varepsilon _{-}+a_{2}\varepsilon _{-}^{2} &=&\left( 2\gamma
-1\right) ^{2}+\left( 2\gamma -1\right) \left( \left( 2\gamma -1\right)
^{2}-2\gamma \right) a_{1}\varepsilon _{-} \\
&&+\varepsilon _{-}^{2}\left( \gamma ^{2}a_{1}^{2}+2\gamma \left( 2\gamma
-1\right) \left( a_{2}-\frac{a_{1}^{2}}{4}\right) -3\left( 2\gamma -1\right)
^{2}\gamma a_{1}^{2}+a_{2}\left( 2\gamma -1\right) ^{4}\right)
\end{eqnarray*}
Identifying order by order the coefficients in each side we then obtain :
\begin{equation}
\left\{
\begin{array}{c}
\gamma =\frac{\omega \Phi _{C}}{\pi }=1 \\
0.a_{1}=0 \\
a_{2}=\frac{5}{4}a_{1}^{2}
\end{array}
\right.
\end{equation}
The first of these equations simply\ translates the fact that, in the small
oscillations limit, $x(\varphi )$ executes harmonic oscillations, which
period is :
\begin{equation}
2\Phi _{C}=\frac{2\pi }{\sqrt{\frac{W_{L}^{\prime \prime }(x_{0})}{m}}}
\end{equation}
The second equation tells us any information. As to the last, it connects
the third and fourth derivatives of the Clairaut's potential in $x_{0}$ :
\begin{equation}
W_{L}^{\left( 4\right) }(x_{0})=\frac{5}{3}\frac{\left( W_{L}^{\left(
3\right) }(x_{0})\right) ^{2}}{m\omega ^{2}}
\end{equation}

Of course, these constraints on the Clairaut's potential, generated by the
isochronism condition, are exactly the same that those obtained in a
singular perturbation expansion \cite{zarmi}, due to the suppression of the
corrections on frequency.

Therefore to generate isochronous oscillations near $x_{0}$ (period $2\Phi
_{C}=\frac{2\pi }{\sqrt{\frac{W_{L}^{\prime \prime }(x_{0})}{m}}}$), the
Clairaut's potential has to satisfy this constraint.

If we refer to the class of potentials $W_{\pm \nu}(x)$ (\ref{potClairaut2}),
this gives ($W_{\pm \nu}^{\left( 4\right) }(x_{0})=\frac{\left( \mp \nu-3\right)
}{x_{0}}W_{\pm \nu}^{\left( 3\right) }(x_{0}),\ W_{\pm \nu}^{\prime \prime
}(x_{0})=m+\nu\left( \nu\pm 1\right) A_{\pm }x_{0}^{\mp \nu-2}=m\left( 2\pm \nu
\right) $) :
\begin{equation}
W_{\pm \nu}^{\left( 3\right) }(x_{0})\left( \frac{\pm \nu+3}{x_{0}}+\frac{5}{3}%
\frac{W_{\pm \nu}^{\left( 3\right) }(x_{0})}{m\left( 2\pm \nu \right) }\right)
=0
\end{equation}
For a potential $W_{\nu}(x),\ \nu >0$, we obtain :
\begin{equation}
W_{\nu}^{\left( 3\right) }(x_{0})\frac{2\left( 2-\nu\right) }{3x_{0}}=0
\end{equation}
which implies $\nu=2$.

For a potential $W_{-\nu}(x),\ 0<\nu<2$, we have on the other hand :
\begin{equation}
W_{-\nu}^{\left( 3\right) }(x_{0})\frac{2\left( \nu+2\right) }{3x_{0}}=0
\end{equation}
which implies $W_{-\nu}^{\left( 3\right) }(x_{0})=0$ that is $\nu =1$.

We recover the preceding result. Nevertheless, it has required much more
tedious calculations, which, as we already pointed out, is an inherent
feature of the perturbative schemes of proof \cite{tikoch},\cite{gold},\cite{brown},\cite{zarmi}.

\end{document}